\documentclass[fleqn,usenatbib]{mnras}

\usepackage{newtxtext,newtxmath}
\usepackage[T1]{fontenc}
\usepackage{ae,aecompl}

\usepackage{graphicx}
\usepackage{amsmath}
\usepackage{amsfonts}
\usepackage{amssymb}
\usepackage{hyperref}
\usepackage{color}

\newcommand{\astroph}[1]{\href{http://arxiv.org/abs/#1}{arXiv:astro-ph/{#1}}}
\newcommand{\arxiv}[1]{\href{http://arxiv.org/abs/#1}{arXiv:{#1}}}
\newcommand{\email}[1]{\href{mailto:#1}{#1}}

\newcommand{\revision}[1]{{#1}}
\newcommand{\revisionb}[1]{{#1}}
\newcommand{\tzero}{{\ensuremath{t_0}}}
\newcommand{\nickel}{{\ensuremath{^{56}\mathrm{Ni}}}}
\newcommand{\cobalt}{{\ensuremath{^{56}\mathrm{Co}}}}
\newcommand{\Mch}{\ensuremath{M_{\rm Ch}}}
\newcommand{\Mni}{\ensuremath{M_{\rm Ni}}}
\newcommand{\Ek}{\ensuremath{E_{\rm k}}}
\newcommand{\En}{\ensuremath{E_{\rm n}}}
\newcommand{\Eb}{\ensuremath{E_{\rm b}}}
\newcommand{\ve}{\ensuremath{v_{\rm e}}}
\newcommand{\vt}{\ensuremath{v_{\rm t}}}
\newcommand{\Msun}{{~\rm M_{\sun}}}
\newcommand{\erg}{{~\rm erg}}
\newcommand{\amu}{{~\rm amu}}
\newcommand{\days}{{~\rm days}}
\newcommand{\kms}{{~\rm km\,s^{-1}}}
\newcommand{\gcm}{{~\rm g\,cm^{-2}}}

\title[SN~Ia asymmetry and $t_0$]{Constraining Type Ia supernova asymmetry with the gamma-ray escape timescale}

\author[Levanon \& Soker]{
Naveh Levanon$^{1}$\thanks{\email{nlevanon@campus.technion.ac.il}, \email{soker@physics.technion.ac.il}}
and Noam Soker$^{1,2}$
\\
$^{1}$Department of Physics, Technion -- Israel Institute of Technology, Haifa 32000 Israel\\
$^{2}$Guangdong Technion Israel Institute of Technology, Shantou 515069, Guangdong Province, China
}

\date{Accepted XXX. Received YYY; in original form ZZZ}

\pubyear{2019}

\begin{document}
\label{firstpage}
\pagerange{\pageref{firstpage}--\pageref{lastpage}}
\maketitle

\begin{abstract}
    We calculate the effects of an asymmetric $^{56}\mathrm{Ni}$ distribution in Type Ia supernova (SN~Ia) ejecta on the \revision{gamma-ray escape timescale ($t_0$) that characterizes the late light curve ($>40\days$ after peak) and find the effect is modest compared to other possible variations in ejecta structure.}
    We parameterize asymmetry in the $^{56}\mathrm{Ni}$ distribution and calculate $t_0$ for a grid of SN ejecta models spanning a large volume of the asymmetry parameter space.
    The models have spherical density profiles while the $^{56}\mathrm{Ni}$ distribution in them has various levels of asymmetry.
    By placing constraints based on the observational measurement of $t_0$ and other general properties of SN~Ia ejecta, we find the range of allowed asymmetry in the $^{56}\mathrm{Ni}$ distribution.
    \revisionb{Considering these constraints} \revision{we find that some level of asymmetry in the distribution is not ruled out.
    However,} models with a single ejecta mass and varying $^{56}\mathrm{Ni}$ distributions cannot explain the full range of observed $t_0$ values.
    This strengthens the claim that both Chandrasekhar mass and sub-Chandrasekhar mass explosions are required to explain the diversity of SN~Ia observations.
\end{abstract}


\section{Introduction}
\label{sec:Introduction}

There are five different binary scenarios whose supporters claim can account for a large fraction, or even all, of type Ia supernovae (SNe~Ia).
These scenarios evolve a white dwarf (WD) or two WDs to explode as a SN~Ia.
The exploding WD can have a mass very close to the Chandrasekhar limit (\Mch\ scenarios) or be of a lower mass, (sub-\Mch\ scenarios).
Each scenario might have two or more channels, with some overlaps between the scenarios in some processes and channels
(for recent reviews on these five scenarios that include many references to earlier papers and reviews see \citealt{LivioMazzali2018, Soker2018Rev, Wang2018, RuizLapuente2019}). 
The five scenarios are as follows.
The core degenerate (CD) scenario (e.g., \citealt{Kashi2011, TsebrenkoSoker2015}; \Mch), 
the double degenerate (DD) scenario (e.g., \citealt{Webbink1984, Zenatietal2019}; sub-\Mch), 
the double-detonation (DDet) scenario (e.g., \citealt{Livne1995, Shenetal2018}; sub-\Mch), 
the single degenerate (SD) scenario (e.g., \citealt{Whelan1973, Wuetal2016}; \Mch), 
and the WD-WD collision (WWC) scenario (e.g., \citealt{LorenAguilar2010, Kushniretal2013}; sub-\Mch). 

There are several key observations that constrain one or more scenarios, and there is no scenario that is free of drawbacks.
This has brought many researchers to conclude that two or more scenarios account for SNe Ia.
In particular, many conclude that one scenario should be a sub-\Mch\ scenario and one must be an \Mch\ scenario.
\revision{This conclusion reflects on several properties of SNe~Ia, e.g., that brighter SNe~Ia have slower light curves \citep[the width-luminosity relation;  e.g.,][]{Kasen2007,Sim2010,Wygoda2019a,Wygoda2019b}.}

Recently more studies focused on late time observations taken $>20 \days$ after light curve peak.
At these times the ejecta becomes transparent and its emission depends on its global structure.
Some studies looked at late-time spectra for clues of ejecta composition \citep[e.g.,][]{Childress2015,Maguire2018,Floers2018}.
Others utilized the late light curves to constrain the ejecta properties \citep[e.g.,][]{Stritzinger2006,Scalzo2014a,Scalzo2019,Wygoda2019a}.
Our work is in the vein of these last studies and attempts to constrain ejecta asymmetry.

There are multiple indications that SN~Ia explosions have some degree of asymmetry.
The polarization measured for SNe~Ia is small but not zero \citep[e.g.,][]{Bulla2016,Yang2019} hinting that some level of asymmetry exists.
\citet{Maeda2010} found shifts in nebular spectra features and explain them using an asymmetric distribution of iron group elements (IGEs).
\citet{Dong2018} found line shifts indicating an off-center \nickel\ distribution in a sub-luminous event.
\citet{Black2019} found narrow absorption features that may indicate IGE clumps are present at high velocities.
Finally, SN remnants (SNRs) display an overall spherical shape yet exhibit clumping \citep[e.g.,][]{Li2017,Sato2019} and small axisymmetrical morphological features \citep[e.g.,][]{TsebrenkoSoker2013}.

Models for SNe~Ia predict a variety of asymmetric structures.
Some progenitor scenarios inherently include an explosion in a non-spherical configuration, such as in a violent merger \citep{Pakmor2012} or a WWC scenario \citep[e.g.,][]{Dong2015}.
The DDet scenario can also include an off-center explosion and a non-spherical \nickel\ distribution due to the geometry of the two detonations \citep[e.g.,][]{Moll2013}.
Central explosions also exhibit asymmetry in the form of clumping in 3D simulations \citep[e.g.,][]{Seitenzahl2013}, though this does not produce pronounced asymmetry.

This paper explores how the late-time light curve constrains possible asymmetry in the \nickel\ distribution in SNe~Ia.
We do this by relying on the gamma-ray escape timescale \tzero\, which is a useful measure for the global ejecta properties including the \nickel\ distribution.
In section \ref{sec:method} we explain how \tzero\ relates to the late time light curves and how we calculate it in our models.
In section \ref{sec:constraints} we list additional constraints we place on the models to find which are potentially viable.
We present and discuss our results in section \ref{sec:results} and summarize them in section \ref{sec:summary}.

\section{Asymmetric models and calculating \tzero}
\label{sec:method}

The definition of the gamma-ray escape timescale \tzero\ follows \citet{Jeffery1999}.
At early times all gamma-rays produced in \nickel\ and \cobalt\ decay are thermalized in the thick ejecta and all of the decay energy is deposited in the ejecta (except for neutrino energy which escapes).
At later times ($>40 \days$) \revision{over 99 per cent of} \nickel\ has decayed so that the dominant energy source is \cobalt.
The ejecta becomes optically thin to gamma-rays and only a fraction of their energy is deposited.
We approximate the \cobalt\ gamma-ray opacity as a grey opacity with $\kappa=0.025\gcm$ \citep{Swartz1995}.
\revision{We assume homologous expansion.}
The average optical depth to \cobalt\ gamma-rays behaves as $\bar\tau = (\tzero/t)^2$ and is unity at the
gamma-ray escape timescale \tzero.
The deposition fraction is then $1 - e^{-(\tzero/t)^2}$.
At this phase the ejecta is also transparent in optical wavelengths so the deposited energy is emitted immediately \revision{\citep{Pinto2000}}.
The bolometric luminosity is therefore
\begin{multline}
    L_{\rm bol}\left(t\right) =
    \frac{N_{\rm Ni}}{t_{\rm Co} - t_{\rm Ni}} 
    \left( e^{-t/t_{\rm Co}} - e^{-t/t_{\rm Ni}} \right) \\
    \times \left[ Q_{{\rm Co},\gamma} \left( 1 - e^{-\left(\tzero/t\right)^2} \right) + 
    Q_{\rm Co,pos} \right],
\label{eq:bolometric luminosity}
\end{multline}
where $N_{\rm Ni}=\Mni/56\amu$ is the initial number of \nickel\ atoms, $t_{\rm Ni}$ and $t_{\rm Co}$ are \nickel\ and \cobalt\ e-folding times, respectively, and 
$Q_{{\rm Co},\gamma}$ and $Q_{\rm Co,pos}$ are 
\cobalt\ gamma-ray and \cobalt\ positron decay energies, respectively.

\revision{
Equation \ref{eq:bolometric luminosity} is an approximation suitable for the light curve $>40$ days after explosion.
While we do not use this equation in this work to infer \tzero\ for individual SNe, it is worthwhile to briefly discuss its accuracy.
A more comprehensive discussion can be found in \citet{Jeffery1999}.
}

\revision{
Deviations from homologous expansion can reach about 10 per cent in the density profile \citep[e.g.,][]{Pinto2000,Woosley2007,Noebauer2017}.
However these deviations occur at early times when the ejecta is still optically thick, and at late times homologous expansion is an excellent approximation.
The ejecta structures we discuss should therefore be considered as the structures after any radiation-driven expansion post explosion.
}

\revision{
We neglect the contribution of \nickel\ to energy deposition because most of it has decayed.
At 40 days past explosion the contribution of \nickel\ to energy deposition if it was fully trapped is about 15 per cent and drops exponentially for later times.
The actual contribution should be less than half this amount as most  \nickel\ gamma-rays escape the ejecta like \cobalt\ gamma-rays do.
\revision{The prescription $1 - e^{-(\tzero/t)^2}$ for the deposition fraction is accurate up to a few per cent \citep{Sukhbold2019}.}
Calculating these errors for the deposition function gives systematic errors in $L_{\rm bol}(t)$ of about 10 per cent at $t=40 \days$ and dropping afterwards.
This induces an error of up to 2.5 days in estimating \tzero\ from $L_{\rm bol}(t)$.
}

There are several methods to estimate the \nickel\ mass $\Mni$ from observations.
The most common method is to relate \Mni\ to the light curve peak using Arnett's rule \citep{Arnett1982}.
Other methods include using calibrated analytic models \citep{Khatami2018} and integrating the luminosity up to late times \citep{Katz2013,Wygoda2019a} which avoids making assumptions about the relation between deposited and emitted energy at the light curve peak.

Fitting equation \ref{eq:bolometric luminosity} requires high quality multi-band light curves from which a quasi-bolometric light curve can be constructed.
Such quasi-bolometric light curves were constructed for several dozen SNe~Ia \citep{Stritzinger2006,Scalzo2014a,Scalzo2019}, and the values of \tzero\ from fitting these light curves are in the range of 30-45 days \citep{Wygoda2019a}.
\revision{We use this range to constrain our models in section \ref{sec:results}.
The \tzero\ values derived from observations suffer from the errors in equation \ref{eq:bolometric luminosity} mentioned above, as well as from observational errors.
\citet{Scalzo2014a} fit \tzero\ in a Bayesian framework and obtained errors of 4-6 days for individual SNe at a 68 per cent confidence level.
\citet{Wygoda2019a} infer \tzero\ in a different method and obtained values similar to \citet{Scalzo2014a} with an average absolute difference of 2 days.
\citet{Papadogiannakis2019} infer a \tzero\ range of 26-40 days for a different set of SNe.
This is consistent with the ranges found by \citet{Scalzo2014a} and \citet{Wygoda2019a} considering the errors in deriving \tzero.
A range of \tzero\ values spanning about 15 days appears robust between these studies.}

The range of observed \tzero\ values constrains the SN~Ia ejecta geometry, as \tzero\ depends on the density profile and the \nickel\ distribution within the ejecta.
With the existing assumptions the average optical depth for \cobalt\ gamma-rays at late times is
\begin{equation}
    \bar\tau =
    \frac{\int d^3 v \, \rho_0 (\vec v) X_{\rm Ni} (\vec v) 
          \oint \frac{d\Omega}{4\pi} \tau ({\vec v}, {\vec \Omega})}
          {\int d^3 v \, \rho_0 (\vec v) X_{\rm Ni} (\vec v)}
\label{eq:average tau}
\end{equation}
where $\rho_0 (\vec v)$ is the density at time \tzero\ and $X_{\rm Ni} (\vec v)$ is the initial \nickel\ distribution after nucleosynthesis.
In equation \ref{eq:average tau} we average $\tau$ over all locations and directions and weigh it by the \nickel\ distribution.
The optical depth for a given location $\vec v$ and direction $\vec \Omega$ is
\begin{equation}
    \tau ({\vec v}, {\vec \Omega}) = \left(\frac{\tzero}{t}\right)^{2}
           \int_0^{\infty} dv_s \, \tzero \kappa \rho_0 ({\vec v} + v_s {\vec \Omega}).
\label{eq:tau}
\end{equation}
Equivalently for a constant opacity we can take $\kappa$ out of the integrals and calculate an average column density \citep{Wygoda2019a}.
For a model with given $\rho_0 (\vec v)$ and $X_{\rm Ni} (\vec v)$, setting $\bar\tau (\tzero) = 1$ gives an expression for $\tzero$ that can be numerically calculated.

To explore the quantitative effect of \nickel\ distribution asymmetry on \tzero\ we define sets of toy models and compute their \tzero\ values.
We use spherical density profiles with either an exponential or a broken power-law shape.
For the exponential density profile we take \citep{Dwarkadas1998}
\begin{equation}
    \rho_0(v) = \frac{M}{8 \pi \left( \ve \tzero \right)^3} e^{-v/\ve}, \quad
    \ve = \left( \frac{\Ek}{6M} \right)^{1/2},
\label{eq:exponential profile}
\end{equation}
where $M$ and \Ek\ are the total ejecta mass and kinetic energy, respectively.
For this profile \citet{Jeffery1999} obtains
\begin{equation}
    \tzero = \sqrt{\frac{M \kappa q}{8 \pi}}\frac{1}{v_e}
        = \sqrt{\frac{3 M^2 \kappa q}{4 \pi \Ek}},
\label{eq:tzero for exponential}
\end{equation}
where $q$ is a unitless structure parameter derived from equations \ref{eq:average tau} and \ref{eq:tau} after inserting the density profile and taking all parameters out.
The value of $q$ depends on the distribution of \nickel\ and on the density profile only and is independent of total mass or time.
It varies between 1 for all \nickel\ in the center and 1/3 for \nickel\ distributed evenly (in mass fraction) in the ejecta.

For the broken power-law profile we take \citep{Chevalier1989}
\begin{equation}
    \rho_0(v) = \frac{\zeta_\rho M}{\left(\vt \tzero \right)^3} 
        \begin{cases}
            \left( v / \vt \right)^{-\delta}, &\text{if} \, v<\vt \\
            \left( v / \vt \right)^{-n}, &\text{otherwise}
        \end{cases}, \quad
    \vt = \zeta_v \left( \frac{\Ek}{M} \right)^{1/2},
\label{eq:powerlaw profile}
\end{equation}
with $\delta$ and $n$ the power-law indices in the inner and outer regions separated by the transition velocity \vt.
The constants $\zeta_\rho$ and $\zeta_v$ are
\begin{equation}
    \zeta_\rho = \frac{\left(3-\delta\right)\left(n-3\right)}
                      {4 \pi \left(n-\delta\right)} , \quad
    \zeta_v = \left[ \frac{2\left(5-\delta\right)\left(n-5\right)}
                          {\left(3-\delta\right)\left(n-3\right)} \right]^{1/2}.
\end{equation}
In a similar manner we obtain for this profile
\begin{equation}
    \tzero = \sqrt{\zeta_\rho M \kappa q^\prime}\frac{1}{v_t}
        = \sqrt{\frac{\zeta_\rho M^2 \kappa q^\prime}{\zeta_v^2 \Ek}},
\label{eq:tzero for powerlaw}
\end{equation}
where a similar structure parameter $q^\prime$ again governs the geometrical contribution to \tzero.

For both spherical profiles we introduce asymmetry to the \nickel\ distribution as follows.
We place \nickel\ in a spherical IGE region of radius $v_{\rm IGE}$.
We displace the \revision{center of the} IGE region from the ejecta center by a velocity $v_{\rm offset}$.
The \nickel\ fraction $X_{\rm Ni}$ is constant within this region.
\revision{We assume the rest of the material in the IGE region is stable IGEs.}
This toy model is not meant to be representative of any physical ejecta model.
We use this simplified asymmetric model to facilitate a parameter space study of the effect of asymmetry on \tzero.

\section{Additional constraints}
\label{sec:constraints}

Apart from the observed values of \tzero\ we place additional constraints on our models to filter out those that are ruled out by observations or by additional theoretical relations.

\begin{enumerate}
    \item The total \nickel\ mass \Mni\ should be in the range $0.3 - 0.8 \Msun$ appropriate for normal SNe~Ia (e.g., \citealt{Scalzo2014a,Childress2015}).
        Our models do not directly define \Mni\ \revision{since \tzero\ is only dependent on $X_{\rm Ni}$}.
        Instead, they define the total IGE zone mass, and we assume the constant $X_{\rm Ni}$ within that zone is in the range $0.65 - 0.9$ \citep{Krueger2012,Seitenzahl2013}, \revision{which gives a range of \Mni\ values per model.
        We set this constraint to analyze models appropriate for normal SNe~Ia.
        We also discuss the effect of relaxing this constraint in the next section.}
    \item \revision{The total kinetic energy \Ek\ should be in a range dictated by conservation of energy in the thermonuclear explosion.
        \Ek\ depends on \Mni\ through the relation} $\Ek = \En - \Eb$, where \En\ and \Eb\ are nuclear energy and WD binding energy, respectively.
        \En\ depends on \Mni\ and the mass of stable IGEs, intermediate mass elements and unburned CO as in \citet{Maeda2009}.
        We take an unburned CO mass fraction of up to 0.05 as in \citet{Scalzo2014a}.
        \Eb\ is in the range $3 - 6 \times 10^{50} \erg$ \citep{Yoon2005}.
        We do not use the binding energy formula of \citet{Yoon2005} directly as their models have masses $> 1.3 \Msun$ and so for sub-\Mch\ WDs this would mean extrapolating from their models.
        Instead we use the above range for \Eb\ and assign an error of $3 \times 10^{50} \erg$ to \Ek\ that we calculate to account for the uncertainties in both \En\ and \Eb.
        \revision{In this method we obtain a range of allowed \Ek\ values per model, and disqualify models with \Ek\ outside this range.}
    \item Observations of IGE velocities in late-time spectra constrain the maximum IGE zone velocity.
        \citet{Maeda2010} use a \nickel\ zone going up to $12000 \kms$ in their toy model to fit nebular spectra.
        \citet{Mazzali2015} find \nickel\ up to $8500 \kms$ in SN~2011fe, and \citet{Dhawan2018} find \nickel\ up to $8600 \kms$ in SN~2014J.
        \citet{Black2019} show evidence for iron clumps at velocities from $8500$ to $12000 \kms$.
        We therefore take a lenient threshold of $12000 \kms$ for the speed of the shallowest \nickel.
\end{enumerate}

These constraints have intentionally lax limits to rule out clearly nonphysical setups yet still allow models that more rigorous treatment may rule out.
We will show in the next section that these constraints are sufficient to understand the allowed level of asymmetry when combined with the \tzero\ observational constraint.

\section{Results \& Discussion}
\label{sec:results}

\begin{figure*} 
  \includegraphics[width=\textwidth,trim=2cm 1.4cm 3cm 2cm,clip]{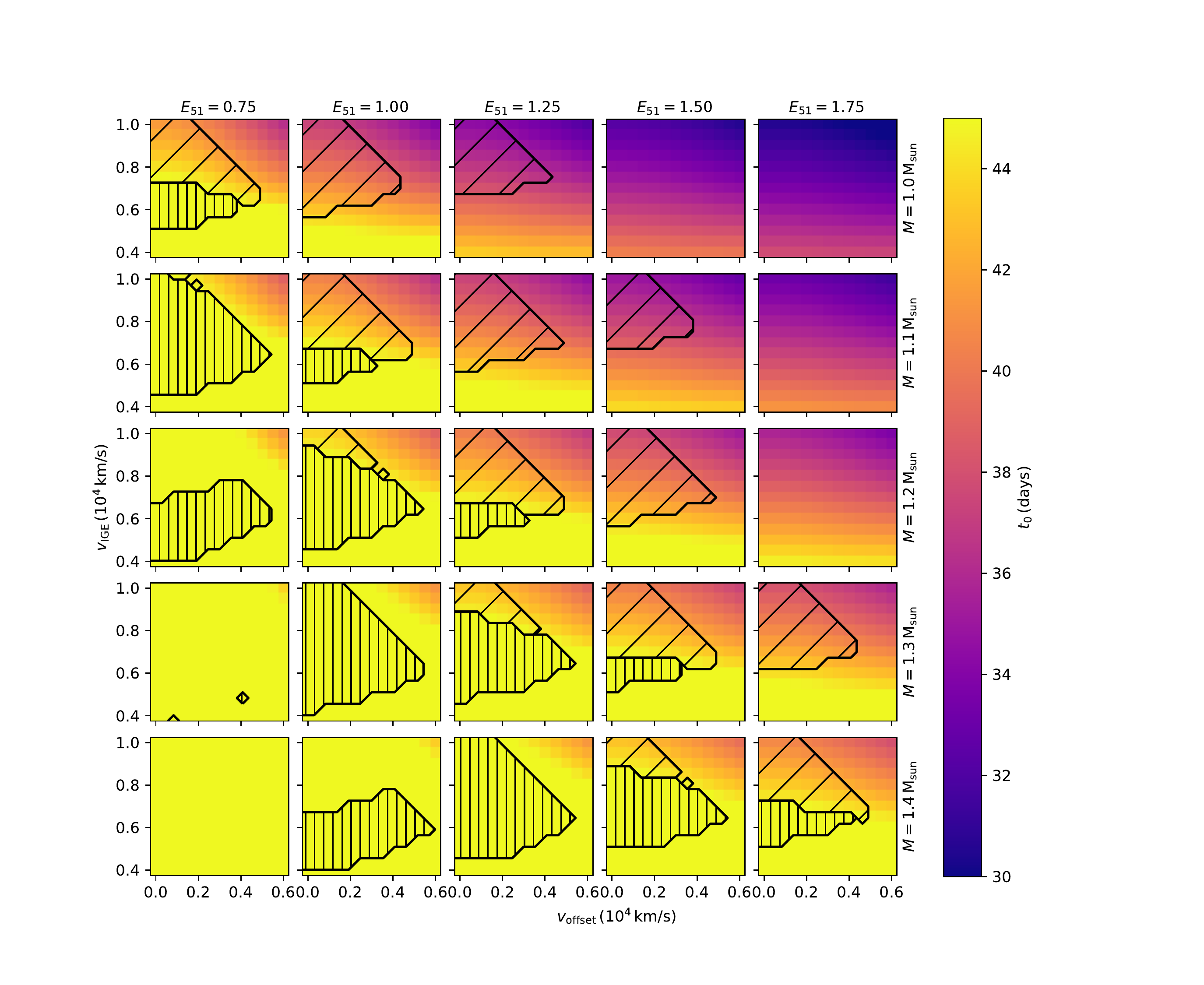}
  \caption{The gamma-ray escape timescale \tzero\ for \revision{models with exponential density profiles}.
  The diagonally hatched regions specify models passing all constraints as detailed in section \ref{sec:constraints} with \tzero\ in the range $30-45 \days$.
  The vertically hatched regions specify models passing all constraints except for \revision{$\tzero < 45$}.
  \revision{The panels represent models with different kinetic energy and total mass in the horizontal and vertical directions, respectively.
  Each panel shows \tzero\ values for models with different $v_{\rm offset}$ (x-axis) and $v_{\rm IGE}$ (y-axis) values.}}
  \label{fig:exponential_constraints}
\end{figure*}
We calculated the gamma-ray escape timescale \tzero\ for models with $M = 0.9-1.4 \Msun$ and $\Ek = 0.5-1.75 \times 10^{51} \erg$ with both an exponential profile (equation \ref{eq:exponential profile}) and broken power-law profiles (equation \ref{eq:powerlaw profile}) with $(\delta,n)=(1,10),(0,10),(0,8)$.
For the velocity radius of the IGE zone $v_{\rm IGE}$ and its velocity offset from the ejecta center $v_{\rm offset}$ we used values between 0 and $10000 \kms$.
We present the results in two figures for the two density profiles, each figure with 25 panels covering different values of the above four parameters.

Fig. \ref{fig:exponential_constraints} shows the values of \tzero\ for exponential models.
Each panel includes models with specific $M$ and $\Ek$, where the x-axis is $v_{\rm offset}$ and the y-axis is $v_{\rm IGE}$.
We mark regions of model parameter space where the models fulfill all constraints with diagonal hatches.
Typically these regions are where $v_{\rm IGE}$ is large enough to have a sufficient mass of \nickel,
and $v_{\rm offset}$ is not too large so that $v_{\rm IGE} + v_{\rm offset} < 12000 \kms$.
\tzero\ decreases as \nickel\ spreads out more and as its offset from the center grows.
The leftmost \revision{side} of each panel describes spherical models (i.e. with $v_{\rm offset}=0$) and asphericity grows towards the right, with upper parts having a more pronounced asymmetry with regards to \tzero\ as \nickel\ spreads out.
The vertically hatched regions fulfill all constraints except for \revision{$\tzero < 45$}, illustrating its importance in limiting the model parameter space.

\begin{figure*}
  \includegraphics[width=\textwidth,trim=2cm 1.4cm 3cm 2cm,clip]{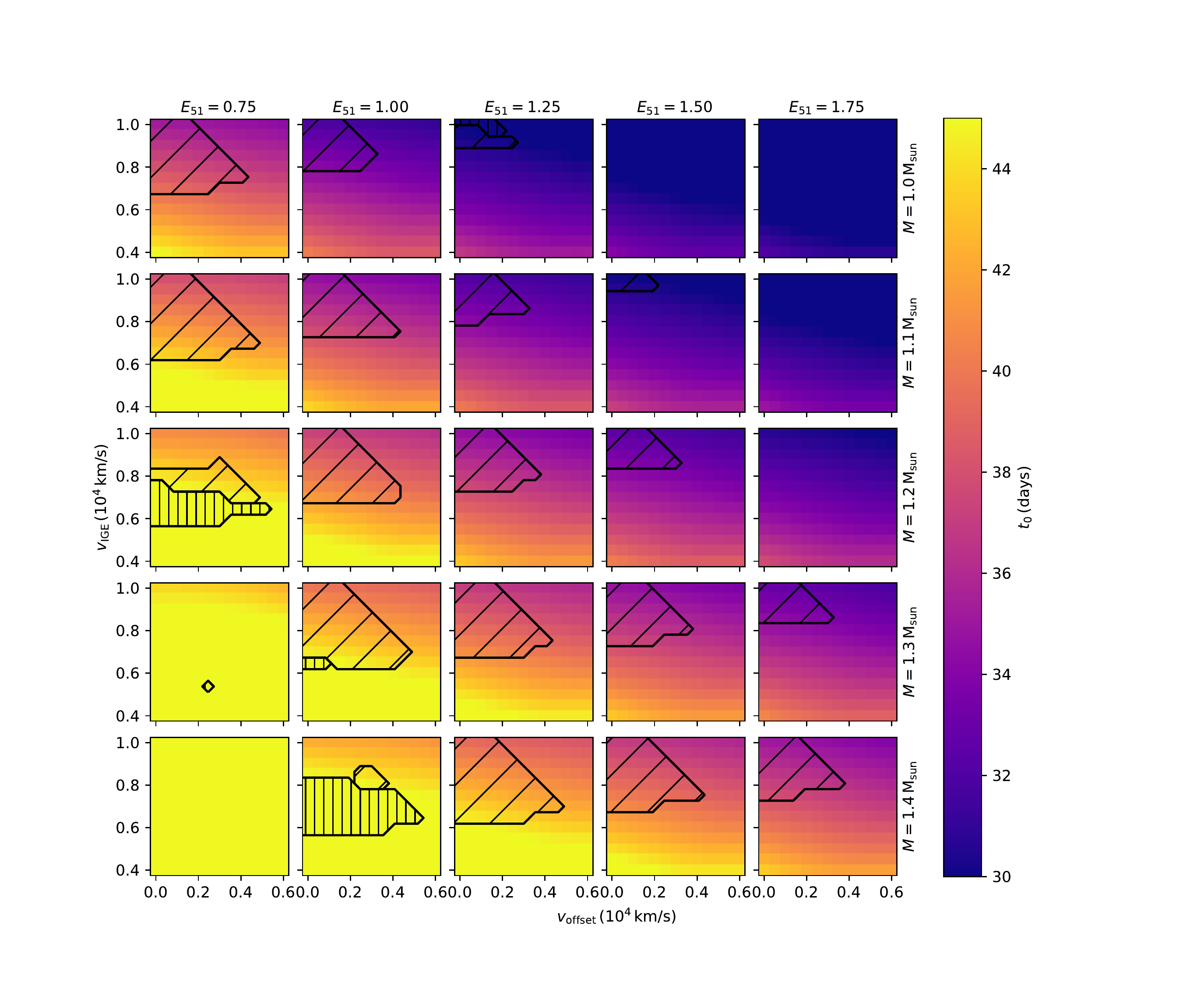}
  \caption{The gamma-ray escape timescale \tzero\ for \revision{models with broken power-law density profiles} with $\delta=1$ and $n=10$.
  Details are the same as for Fig. \ref{fig:exponential_constraints}}
  \label{fig:powerlaw_constraints}
\end{figure*}

Fig. \ref{fig:powerlaw_constraints} shows the same as Fig. \ref{fig:exponential_constraints} for broken power-law profiles with $\delta=1$ and $n=10$.
The power-law density profiles are less centrally steep than the exponential ones, so that optical depths are smaller for given model parameters.

\revision{
The model parameters which are directly inferred from bolometric light curves are \Mni\ and \tzero.
Fig. \ref{fig:exponential_mni_vs_t0} and Fig. \ref{fig:powerlaw_mni_vs_t0} show these parameters for models that pass all our constraints for the exponential and broken power-law density profiles, respectively.
Different markers correspond to different ejecta masses.
Each model has a range of \Mni\ values between 0.65 and 0.9 of the IGE region mass, and the plotted value is the center of this range.
This is why some models have $\Mni < 0.3 \Msun$ in the plot despite our constraint on \Mni.
For a given total mass $M$, models are roughly grouped by \Ek\ where small \tzero\ correspond to large \Ek\ and vice versa.
For given $M$ and \Ek, models with larger $v_{\rm IGE}$ have larger \Mni\ and smaller \tzero, and models with larger $v_{\rm offset}$ have smaller \Mni\ and \tzero.
These figures illustrate for given $M$ and \Mni\ what values of \tzero\ are expected.
They also show that the variety in \tzero\ induced by the \revisionb{\nickel\ distribution} parameters for given $M$ and \Ek\ is about 2 days for varying $v_{\rm offset}$ and 4 days for varying $v_{\rm IGE}$.
\begin{figure}
    \centering
    \includegraphics[width=\linewidth,trim=0.5cm 0.4cm 1cm 1cm,clip]{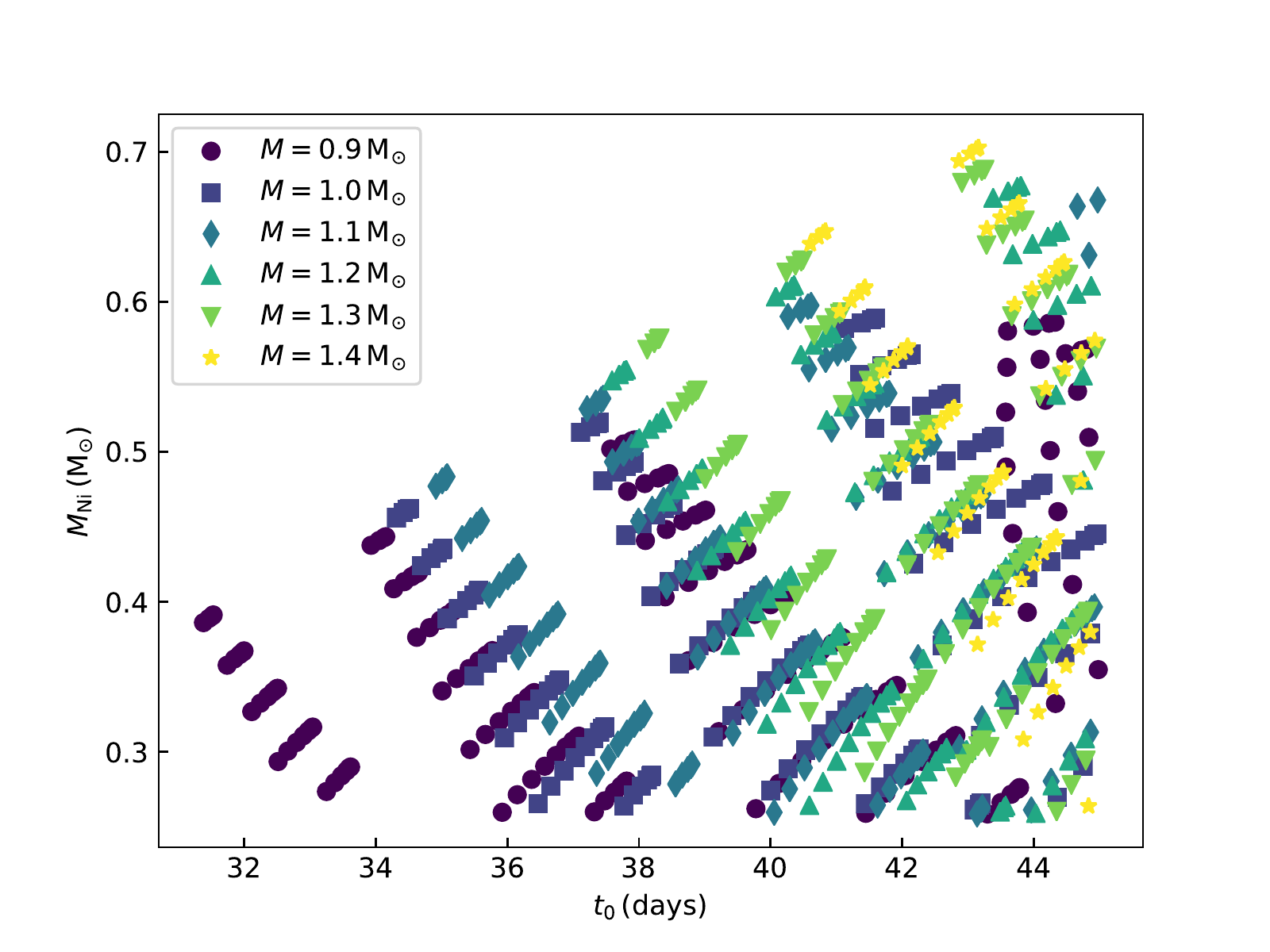}
    \caption{\tzero\ and \Mni\ for models with an exponential density profile passing the constraints in section \ref{sec:constraints}.
    These are the models appearing in the diagonally hatched regions in Fig. \ref{fig:exponential_constraints}.
    Different markers correspond to different total ejecta masses.
    We include models with $M=0.9\Msun$ that were not displayed in Fig. \ref{fig:exponential_constraints}.}
    \label{fig:exponential_mni_vs_t0}
\end{figure}
\begin{figure}
    \centering
    \includegraphics[width=\linewidth,trim=0.5cm 0.4cm 1cm 1cm,clip]{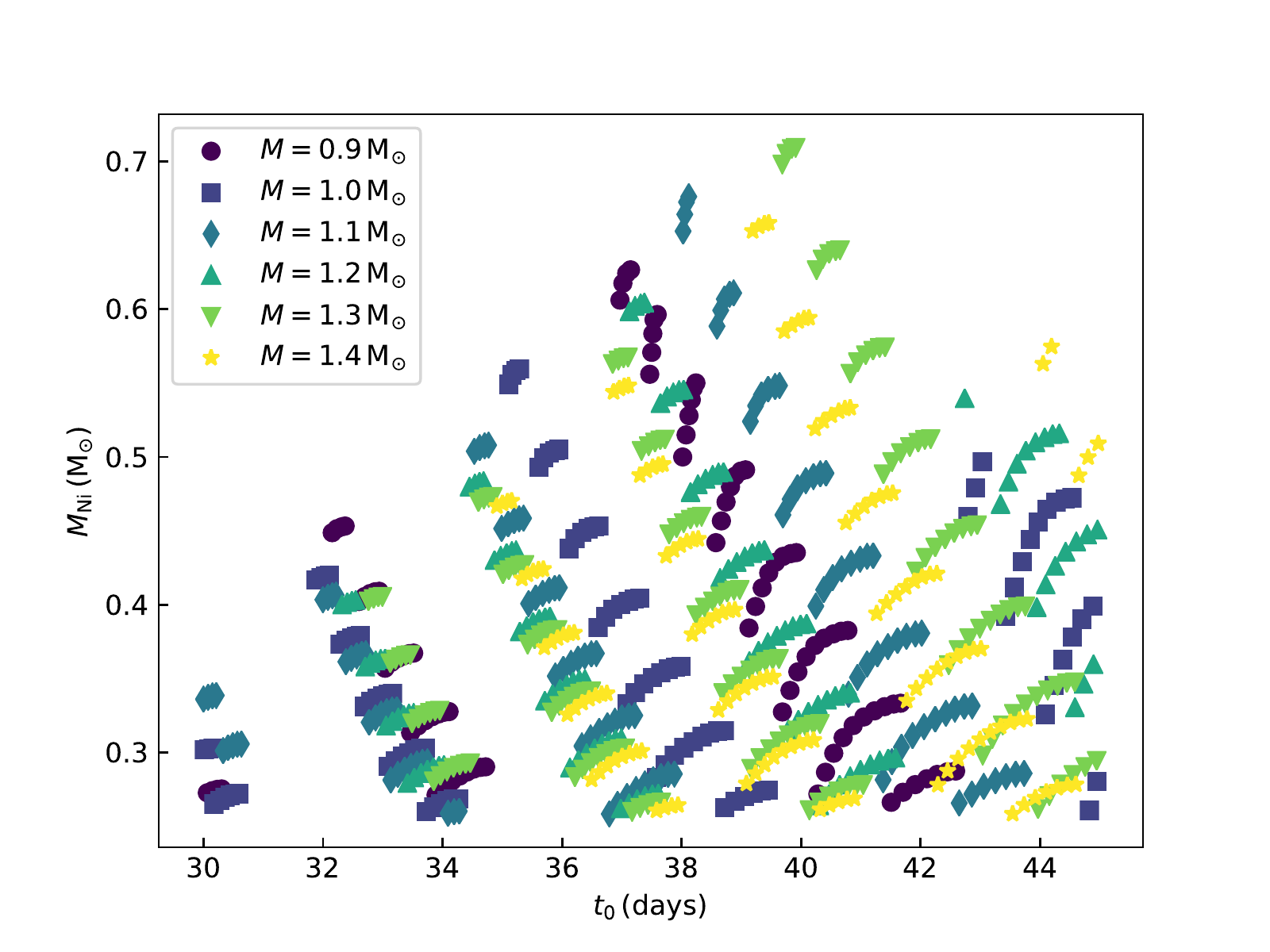}
    \caption{Like Fig. \ref{fig:exponential_mni_vs_t0} for models with a broken power-law density profile.}
    \label{fig:powerlaw_mni_vs_t0}
\end{figure}
}

How much is the \tzero\ constraint constraining?
It limits \nickel\ from being more concentrated in the center with smaller $v_{\rm IGE}$, \revision{or} otherwise \tzero\ \revision{would go} above 45 days.
Without the \tzero\ constraint \nickel\ can be buried deeper in models with an exponential profile.
Models with a broken-power law profile have overall lower inner densities so that \tzero\ is less constraining as seen in the minor vertical hatched regions in Fig. \ref{fig:powerlaw_constraints}.
\citet{Papadogiannakis2019} infer \tzero\ values ranging down to 26 days.
However, relaxing the lower limit of the \tzero\ constraint does not change the results here since it is the other constraints that limit models with small \tzero.

We find that exponential profile \Mch\ models can give \tzero\ values only above 40 days, and so these models cannot explain most of the \tzero\ range.
However, with less steep power-law models we obtain \tzero\ values down to 35 days.
This still does not explain the fast-declining end of the normal SNe~Ia variety, where \tzero\ is 30-35 days.
This is in line with other studies that found that fast-declining normal SNe~Ia cannot be explained by \Mch\ models (e.g. \citealt{Scalzo2014b,Blondin2017,Goldstein2018,Wygoda2019a}).
Our conclusion is stronger even, because our toy models include less assumptions and are therefore more general.

\revision{The general nature of our toy models raises the question of whether more detailed models would change the conclusion regarding \Mch\ models having $\tzero > 35 \days$.
We find that the most significant parameter affecting \tzero\ is the total mass $M$, and the effect of the \nickel\ distribution asymmetry is secondary to it.
\revisionb{This is expected since \tzero\ depends on $M$ more strongly than on $q$ in equations (\ref{eq:tzero for exponential}) and (\ref{eq:tzero for powerlaw}).}
Additionally, we chose weak constraints that allow a greater range of model parameters for given $M$ than would appear in more realistic models.
Considering this we find the result that \Mch\ models are still limited to $\tzero > 35 \days$ compelling.
A different caveat with this result is the uncertainties in the observed range of \tzero\ values we discussed in Section \ref{sec:method}.
If in effect observed \tzero\ values have a minimum of 35 rather than 30 days, then \Mch\ models adequately explain this range.
This cannot be ruled out, though \citet{Papadogiannakis2019} find a range of \tzero\ values extending down to 26 days which is well beyond the reach of \Mch\ models with the current uncertainties.}


The spherical models ($v_{\rm offset}=0$) have the full range of \tzero\ values for models with given $M$ and $\Ek$.
This means asymmetry cannot be invoked to explain lower \tzero\ values once other constraints are accounted for.
One cannot claim a model with too large \tzero\ would in effect have a lower value once we account for a smooth asymmetric \nickel\ distribution as we have here.
We did not model surface \nickel\ or shallow \nickel\ clumps separate from the central blob.
These could potentially influence \tzero\, though their effects on the early light curve may place strong constraints on them \citep[e.g.,][]{Noebauer2017}.
The weak effect of asymmetry on possible \tzero\ values may be because we placed a relatively relaxed constraint on the relation between \Ek\ and \Mni.
We also tested a tightened constraint by using a stricter \Eb\ range of $4-5 \times 10^{50} \erg$ and assigning zero error to the calculated \Ek.
The resulting allowed regions are much smaller and still the difference in \tzero\ ranges between spherical models and asymmetric models is less than half a day.

\revision{
Constraint (i) in Section \ref{sec:constraints} limits the allowed parameter space to models with $0.3 - 0.8 \Msun$ of \nickel\ as appropriate for normal SNe~Ia.
However, sub-luminous SNe~Ia can have lower \nickel\ masses down to $0.05 \Msun$ \citep[e.g.,][]{Stritzinger2005,Dhawan2017}.
Removing the lower bound on \Mni\ allows for models with smaller $v_{\rm IGE}$ and larger $v_{\rm offset}$ extending to the lower part of the panels in Fig. \ref{fig:exponential_constraints} and \ref{fig:powerlaw_constraints}.
However these models have relatively large \tzero\ values.
If we also consider that sub-luminous events mostly have $\tzero < 35 \days$ \citep{Wygoda2019a} we find as above that sub-\Mch\ models are required and that the allowed \revisionb{\nickel\ distribution} parameters are within the regions shown in Fig. \ref{fig:exponential_constraints} and \ref{fig:powerlaw_constraints}.
}

\section{Summary}
\label{sec:summary}

In this paper we explored the effect of an asymmetric \nickel\ distribution on the gamma-ray escape timescale \tzero\ using toy models.
Our approach enabled us to test global asymmetries in the \nickel\ distribution over the entire allowed parameter space.
We limited the asymmetry parameter space using the range of observed \tzero\ values and additional model-independent constraints.
We found that \revisionb{including these additional constraints limits the allowed \nickel\ distribution asymmetry.
The allowed asymmetry includes an IGE region offset from the center by up to $4000 \kms$.
This asymmetry then} does not influence the value of \tzero\ by much, which means it does not impact the decline of the bolometric light curve.
From the opposite perspective this means that we can include a certain level of global \nickel\ asymmetry without affecting the bolometric light curve.
\revisionb{We computed the \tzero\ values corresponding to these allowed asymmetries for a range of \Mch\ and sub-\Mch\ models.}
We conclude that while the bolometric light curve does encode information on the SN ejecta structure in \tzero, it does not enable extracting the level of asymmetry of a given event.
 
While we modelled various \nickel\ distribution asymmetries, the total density distribution remained spherical.
Therefore the expected SNR will be spherical, compatible with observed morphologies of SNe~Ia remnants.
So we turn to address the question of whether the progenitor of SNe~Ia come from sub-\Mch\ scenarios and/or \Mch\ scenarios. 
   
Examining the lower row of both Fig. \ref{fig:exponential_constraints} and Fig. \ref{fig:powerlaw_constraints}, we see that \Mch\ scenarios cannot account for short gamma-ray escape timescales of $\tzero \la 35 \days$.
In the cases of ejecta with exponential density profiles (Fig. \ref{fig:exponential_constraints}) even SNe with masses as low as $1 \Msun$ cannot explain the shortest values of \tzero. 
In the cases of ejecta with broken power-law density profiles (Fig. \ref{fig:powerlaw_constraints}) SNe with masses in the range of $1-1.2 \Msun$ marginally account for the longest values of \tzero, and only with a low explosion energy and a mass of $1.2 \Msun$ in that mass range.
We conclude that the best explanation for our results is that both sub-\Mch\ and \Mch\ WDs explode as SNe~Ia.

\revision{This study cannot resolve the question of which progenitor scenario leads to sub-\Mch-mass explosions and which leads to \Mch-mass explosions, or whether both come from a single progenitor scenario.
However, we prefer that these come from separate progenitor scenarios based on some of our other recent results.}
Our recent results include finding that the DD scenario can account for early excess emission \citep{LevanonSoker2019} and the recent review by \cite{Soker2018Rev} with a table summarizing the five scenarios listed in section \ref{sec:Introduction} which considers many more observational properties of SNe~Ia.
Informed by these studies we take the view that the main sub-\Mch\ scenario is the DD scenario, with possibly some SNe~Ia coming from the DDet scenario, and that the \Mch\ scenario is the CD scenario.

\section*{Acknowledgements}
\revision{We thank an anonymous referee for many helpful comments.}
This research was supported by the Asher Fund for Space Research at the Technion, and the Israel Science Foundation.


\bsp	
\label{lastpage}

\begin{thebibliography}{99}

\bibitem[Arnett(1982)]{Arnett1982}
Arnett, W.~D.\ 1982, \apj, 253, 785 

\bibitem[Black et al.(2019)]{Black2019}
Black, C.~S., Fesen, R.~A., \& Parrent, J.~T.\ 2019, \mnras, 483, 1114.

\bibitem[Blondin et al.(2017)]{Blondin2017} 
\revision{Blondin, S., Dessart, L., Hillier, D.~J., et al.\ 2017, \mnras, 470, 157.}

\bibitem[Bulla et al.(2016)]{Bulla2016}
Bulla, M., Sim, S.~A., Kromer, M., et al.\ 2016, \mnras, 462, 1039.

\bibitem[Chevalier \& Soker(1989)]{Chevalier1989}
Chevalier, R.~A., \& Soker, N.\ 1989, \apj, 341, 867 

\bibitem[Childress et al.(2015)]{Childress2015}
Childress, M.~J., Hillier, D.~J., Seitenzahl, I., et al.\ 2015, \mnras, 454, 3816.

\bibitem[Dhawan et al.(2017)]{Dhawan2017} Dhawan, S., Leibundgut, B., Spyromilio, J., et al.\ 2017, \aap, 602, A118.

\bibitem[Dhawan et al.(2018)]{Dhawan2018}
Dhawan, S., Fl{\"o}rs, A., Leibundgut, B., et al.\ 2018, \aap, 619, A102.

\bibitem[Dong et al.(2015)]{Dong2015}
Dong, S., Katz, B., Kushnir, D., et al.\ 2015, \mnras, 454, L61.

\bibitem[Dong et al.(2018)]{Dong2018}
Dong, S., Katz, B., Kollmeier, J.~A., et al.\ 2018, \mnras, 479, L70.

\bibitem[Dwarkadas \& Chevalier(1998)]{Dwarkadas1998}
Dwarkadas, V.~V., \& Chevalier, R.~A. 1998, \apj, 497, 807.

\bibitem[Fl{\"o}rs et al.(2018)]{Floers2018} 
\revision{Fl{\"o}rs, A., Spyromilio, J., Maguire, K., et al.\ 2018, \aap, 620, A200.}

\bibitem[Goldstein \& Kasen(2018)]{Goldstein2018}
Goldstein, D.~A., \& Kasen, D.\ 2018, \apj, 852, L33.

\bibitem[Jeffery(1999)]{Jeffery1999}
Jeffery, D.~J.\ 1999, arXiv e-prints , \astroph{9907015}

\bibitem[Kasen \& Woosley(2007)]{Kasen2007} 
\revision{Kasen, D., \& Woosley, S.~E.\ 2007, \apj, 656, 661.}

\bibitem[Kashi \& Soker(2011)]{Kashi2011}
Kashi, A., \& Soker, N.\ 2011, \mnras, 417, 1466.

\bibitem[Katz et al.(2013)]{Katz2013}
Katz, B., Kushnir, D., \& Dong, S.\ 2013, arXiv e-prints , \arxiv{1301.6766}.

\bibitem[Khatami \& Kasen(2018)]{Khatami2018}
Khatami, D.~K., \& Kasen, D.~N.\ 2018, arXiv e-prints , \arxiv{1812.06522}.

\bibitem[Krueger et al.(2012)]{Krueger2012}
Krueger, B.~K., Jackson, A.~P., Calder, A.~C., et al.\ 2012, \apj, 757, 175.

\bibitem[Kushnir et al.(2013)]{Kushniretal2013}
Kushnir, D., Katz, B., Dong, S., Livne, E., \& Fern{\'a}ndez, R.\ 2013, \apjl, 778, L37

\bibitem[Levanon \& Soker(2019)]{LevanonSoker2019} 
\revision{Levanon, N., \& Soker, N.\ 2019, \apj, 872, L7.}

\bibitem[Li et al.(2017)]{Li2017}
Li, C.-J., Chu, Y.-H., Gruendl, R.~A., et al.\ 2017, \apj, 836, 85.

\bibitem[Livio \& Mazzali(2018)]{LivioMazzali2018}
Livio, M., \& Mazzali, P.\ 2018, Physics Reports, 736, 1

\bibitem[Livne \& Arnett(1995)]{Livne1995}
Livne, E., \& Arnett, D.\ 1995, \apj, 452, 62.

\bibitem[Lor{\'e}n-Aguilar et al.(2010)]{LorenAguilar2010}
Lor{\'e}n-Aguilar, P., Isern, J., \& Garc{\'{\i}}a-Berro, E.\ 2010, \mnras, 406, 2749

\bibitem[Maeda \& Iwamoto(2009)]{Maeda2009}
Maeda, K., \& Iwamoto, K. 2009, \mnras, 394, 239

\bibitem[Maeda et al.(2010)]{Maeda2010}
Maeda, K., Taubenberger, S., Sollerman, J., et al.\ 2010, \apj, 708, 1703.

\bibitem[Maguire et al.(2018)]{Maguire2018}
Maguire, K., Sim, S.~A., Shingles, L., et al.\ 2018, \mnras, 477, 3567.

\bibitem[Mazzali et al.(2015)]{Mazzali2015}
Mazzali, P.~A., Sullivan, M., Filippenko, A.~V., et al.\ 2015, \mnras, 450, 2631.

\bibitem[Moll \& Woosley(2013)]{Moll2013}
Moll, R., \& Woosley, S.~E.\ 2013, \apj, 774, 137.

\bibitem[Noebauer et al.(2017)]{Noebauer2017}
Noebauer, U.~M., Kromer, M., Taubenberger, S., et al.\ 2017, \mnras, 472, 2787.

\bibitem[Pakmor et al.(2012)]{Pakmor2012}
Pakmor, R., Kromer, M., Taubenberger, S., et al.\ 2012, \apj, 747, L10.

\bibitem[Papadogiannakis et al.(2019)]{Papadogiannakis2019} 
\revision{Papadogiannakis, S., Dhawan, S., Morosin, R., et al.\ 2019, \mnras, 485, 2343.}

\bibitem[Pinto \& Eastman(2000)]{Pinto2000}
\revision{Pinto, P.~A., \& Eastman, R.~G.\ 2000, \apj, 530, 744.}

\bibitem[Ruiz-Lapuente(2019)]{RuizLapuente2019}
Ruiz-Lapuente, P.\ 2019, arXiv e-prints , \arxiv{1812.04977}

\bibitem[Sato et al.(2019)]{Sato2019}
Sato, T., Hughes, J.~P., Williams, B.~J., et al.\ 2019, arXiv e-prints , \arxiv{1903.00764}.

\bibitem[Scalzo et al.(2014)]{Scalzo2014a}
Scalzo, R., Aldering, G., Antilogus, P., et al.\ 2014a, \mnras, 440, 1498.

\bibitem[Scalzo et al.(2014)]{Scalzo2014b}
Scalzo, R.~A., Ruiter, A.~J., \& Sim, S.~A.\ 2014b, \mnras, 445, 2535.

\bibitem[Scalzo et al.(2019)]{Scalzo2019}
Scalzo, R.~A., Parent, E., Burns, C., et al.\ 2019, \mnras, 483, 628.

\bibitem[Seitenzahl et al.(2013)]{Seitenzahl2013}
Seitenzahl, I.~R., Ciaraldi-Schoolmann, F., R{\"o}pke, F.~K., et al.\ 2013, \mnras, 429, 1156.

\bibitem[Shen et al.(2018)]{Shenetal2018}
Shen, K.~J., Kasen, D., Miles, B.~J., \& Townsley, D.~M.\ 2018, \apj, 854, 52

\bibitem[Sim et al.(2010)]{Sim2010} 
\revision{Sim, S.~A., R{\"o}pke, F.~K., Hillebrandt, W., et al.\ 2010, \apj, 714, L52.}

\bibitem[Soker(2018)]{Soker2018Rev}
Soker, N.\ 2018, Science China Physics, Mechanics, and Astronomy, 61, 49502

\bibitem[Stritzinger(2005)]{Stritzinger2005}
Stritzinger, M., 2005, PhD thesis, Technichal Univ. Munich

\bibitem[Stritzinger et al.(2006)]{Stritzinger2006}
Stritzinger, M., Leibundgut, B., Walch, S., et al.\ 2006, \aap, 450, 241.

\bibitem[Sukhbold(2019)]{Sukhbold2019}
\revisionb{Sukhbold, T.\ 2019, \apj, 874, 62.}

\bibitem[Swartz et al.(1995)]{Swartz1995}
Swartz, D.~A., Sutherland, P.~G., \& Harkness, R.~P.\ 1995, \apj, 446, 766.

\bibitem[Tsebrenko \& Soker(2013)]{TsebrenkoSoker2013}
Tsebrenko, D., \& Soker, N.\ 2013, \mnras, 435, 320.

\bibitem[Tsebrenko \& Soker(2015)]{TsebrenkoSoker2015}
Tsebrenko, D., \& Soker, N.\ 2015, \mnras, 447, 2568

\bibitem[Wang(2018)]{Wang2018}
Wang, B.\ 2018, RAA 2018, 18, 49

\bibitem[Webbink(1984)]{Webbink1984}
Webbink, R.~F.\ 1984, \apj, 277, 355.

\bibitem[Whelan \& Iben(1973)]{Whelan1973}
Whelan, J., \& Iben, I.\ 1973, \apj, 186, 1007.

\bibitem[Woosley et al.(2007)]{Woosley2007}
\revision{Woosley, S.~E., Kasen, D., Blinnikov, S., et al.\ 2007, \apj, 662, 487.}

\bibitem[Wu et al.(2016)]{Wuetal2016}
Wu, C.-Y., Liu, D.-D., Zhou, W.-H., \& Wang, B.\ 2016, Research in Astronomy and Astrophysics, 16, 160

\bibitem[Wygoda et al.(2019a)]{Wygoda2019a}
\revision{Wygoda, N., Elbaz, Y., \& Katz, B.\ 2019, \mnras, 484, 3941.}

\bibitem[Wygoda et al.(2019b)]{Wygoda2019b} 
\revision{Wygoda, N., Elbaz, Y., \& Katz, B.\ 2019, \mnras, 484, 3951.}

\bibitem[Yang et al.(2019)]{Yang2019} 
\revision{Yang, Y., Hoeflich, P.~A., Baade, D., et al.\ 2019, arXiv e-prints , \arxiv{1903.10820}.}

\bibitem[Yoon \& Langer(2005)]{Yoon2005}
Yoon, S.-C., \& Langer, N.\ 2005, \aap, 435, 967.

\bibitem[Zenati et al.(2019)]{Zenatietal2019}
Zenati, Y., Toonen, S., \& Perets, H.~B.\ 2019, \mnras, 482, 1135

\end{thebibliography}
\end{document}